\documentclass[11pt]{article}
\pdfoutput=1

\usepackage{amssymb,amsfonts,amsthm,amsmath,amssymb,amscd,amstext,epsfig}
\usepackage{color}

\newcommand{\rhoms}{\tilde{\rho}_{\rm s}}
\newcommand{\rhomn}{\tilde{\rho}_{\rm n}}
\newcommand{\rhom}{\tilde{\rho}}

\newcommand{\bea}{\begin{eqnarray}}
\newcommand{\eea}{\end{eqnarray}}

\newcommand{\ginf}{{\gamma}}

\def\be{\begin{equation}}
\def\ee{\end{equation}}

\begin{document}

\begin{flushright}
PUPT-2305\\
arXiv:0906.4810 [hep-th]
\end{flushright}

\begin{center}
\vspace{1cm} { \LARGE {\bf Sound modes in holographic superfluids}}\\

\vspace{1.1cm}

Christopher P.~Herzog and Amos Yarom

\vspace{0.7cm}

{Department of Physics, Princeton University \\
     Princeton, NJ 08544, USA }

\vspace{0.7cm}

{\tt cpherzog@princeton.edu, ayarom@princeton.edu} \\

\vspace{1.5cm}

\end{center}

\begin{abstract}
\noindent
Superfluids support many different types of sound waves.
We investigate the relation between the sound waves in a relativistic and a non-relativistic superfluid by using hydrodynamics to calculate the various sound speeds. Then, using a particular holographic scalar gravity realization of a strongly interacting superfluid, 
we compute first, second and fourth sound speeds as a function of the temperature.  
The relativistic low temperature results for second sound differ from Landau's well known prediction for the non-relativistic, incompressible case.
\end{abstract}

\pagebreak

\setcounter{page}{1}
\setcounter{equation}{0}

\tableofcontents

\section{Introduction}

In this work, we are interested in the sound speeds of a strongly interacting relativistic superfluid. Since perturbative techniques are not applicable, we use the AdS/CFT correspondence \cite{Maldacena:1997re, Gubser:1998bc, Witten:1998qj} to study the sound modes in these systems.  Relativistic superfluids may be important for understanding the physics of neutron stars  \cite{Carter:2001tg}.
In addition, the holographic model we use has also been employed to model quantum criticality in certain condensed matter systems (see \cite{Hartnoll:2009sz, Herzog:2009xv} for reviews).

A superfluid can be defined as a liquid in which a Bose condensate is formed. 
In a weakly coupled system, a Bose condensed phase occurs whenever 
a large number of particles occupies the ground state of the free theory. 
A more formal description of a Bose condensate is that of a system 
where the single particle density function does not vanish at 
large spatial separation \cite{Penrose:1956}. 
The virtue of the latter definition is that it does not rely 
on a weak-coupling description of the system. 
Indeed many properties of the superfluid phase of ${}^4$He  
can be accounted for when modeled by a strongly-interacting 
Bose condensate. Of particular interest to us in 
this work are the sound modes which may be excited in the superfluid. 

A superfluid at nonzero temperature can be thought of as a two component system: an uncondensed, normal component and a condensed, superfluid component.  Each component
has its own density field and velocity field. The collective motion of the fluid where both components move in phase is called ordinary sound, while second sound is associated with out of phase motion of the two components.
In a non-relativistic and incompressible superfluid one finds that ordinary (or first) sound waves couple more strongly to pressure oscillations while second sound waves are sourced by temperature oscillations \cite{LandLV6}.
There is also a fourth sound mode where the normal component is immobilized, for example by packing a capillary tube with powder. This sound mode propagates through density fluctuations of the superfluid component.\footnote{%
 Third sound is associated with surface waves in a thin film of superfluid \cite{Atkins}.
}
 
Regardless of the coupling strength, a hydrodynamic description of the superfluid enables a calculation of sound speeds from the equation of state and various  
thermodynamic quantities. 
In section \ref{sec:soundmodes}, we describe how such a calculation can be carried out
for the relativistic superfluid and relate the relativistic expressions for the sound modes to the much better known non-relativistic ones. 

The particular 
holographic
model we use is the scalar gravity system introduced by refs.\ \cite{Gubser:2008px, Hartnoll:2008vx} which consists of an Abelian gauge field and a charged scalar field in an electrically charged black hole background with a negative cosmological constant.  
A Higgs mechanism in the bulk gravity theory is dual to spontaneous breaking of a global U(1) symmetry in the boundary field theory.
The authors of ref.\ \cite{Herzog:2008he} 
argued that this system is dual to a strongly interacting, relativistic superfluid.
We give more details regarding the mapping of the gravitational system to the conformal superfluid in section \ref{S:Bulk}.

Our work was motivated in part by
Landau's well known prediction for the low temperature behavior of second sound in a non-relativistic incompressible superfluid \cite{LandLV6}. Landau argued that at low temperatures, 
\be
\label{E:Landaulimit}
\lim_{T\to 0} \tilde c_2^2 = \frac{\tilde c_1^2}{d} \ ,
\ee  
where $\tilde c_i$ is the nonrelativistic speed of $i$'th sound and $d$ is the number of spatial dimensions. 
This argument has been generalized to a relativistic fluid in \cite{Carter:1995if}.
In our setup, we find that the speed of second sound
does not approach this limit. 
In section \ref{S:Numerics} we describe the behavior of second (and fourth) sound for the holographic superfluid. We discuss the difference between \eqref{E:Landaulimit} and the relativistic result in section \ref{S:Discussion}.

Previous work \cite{Herzog:2008he,Basu:2008bh,Herzog:2009ci,Amado:2009ts,Yarom:2009uq} examined the speed of second sound and fourth sound in a probe limit where the scalar and gauge field were not allowed to back react on the metric.  (The probe approximation can also be thought of as a limit where the charge of the scalar field becomes large.)  This limit is not well suited to an investigation of Landau's prediction.  The reason is that in the probe limit the charged matter is a small perturbation of the rest of the system, while Landau's prediction requires an assumption that the whole system become a superfluid at $T=0$.    
In the probe limit fourth sound and second sound coincide and approach first sound at zero temperature provided that the zero temperature limit of the theory is well defined.
In this paper, we do not work in the probe limit and consider the full backreacted geometry.

\section{Sound modes}
\label{sec:soundmodes}

Consider a thermal system with a conserved $U(1)$ symmetry. The hydrodynamic variables used to describe such a state are the local velocity field of the fluid $u^{\nu}$ with $\nu = 0, \ldots, d$, the temperature $T$ and the chemical potential $\mu$. If the $U(1)$ symmetry is spontaneously broken a condensate forms, and the resulting Goldstone boson $\varphi$ provides for a new degree of freedom $\xi^{\nu} = \partial^{\nu} \varphi$.  Refs.\  \cite{Son:2000ht,Herzog:2008he} developed a relativistic hydrodynamic description of such a system in the ideal limit, ignoring dissipation.\footnote{%
See refs.\ \cite{Israelold, KhalatnikovRelativistic, Carterbook, Valle:2007xx} for earlier work on the relativistic hydrodynamics of superfluids.
}
In what follows we summarize their results. 
The extra degree of freedom $\xi^{\nu}$ is interpreted as the (non-normalized) velocity of the condensate $v^{\nu}$ via $\xi^{\nu} = \mu v^{\nu}$. The norm of the condensate velocity, $\beta$, defined through $(1-\beta^2) = -v_{\nu}v^{\nu}$ serves as the new hydrodynamic degree of freedom.
The velocity field $u^{\nu}$ denotes the velocity of the uncondensed phase. It can be defined as the local boost parameter required to bring the superfluid velocity to the form $v^{\mu} = (1,\beta {\bf n})$ where $\bf n$ is a spatial 3-vector of unit norm.
The pressure $P$, being a Lorentz scalar, can depend on $T$, $\mu$ and $\beta$. 
We define the entropy density $s$, normal density $\rho_{\rm n}$ and condensate density $\rho_{\rm s}$ as  the variables conjugate to temperature, chemical potential and the norm of the superfluid four-velocity,
\begin{equation}
\label{E:dP}
	dP = s \, dT +\rho_{\rm n} d\mu - \frac{\rho_{\rm s}}{2\mu} d \xi^2\,,
\end{equation}
where $\xi^2 = \xi_{\nu}\xi^{\nu}$.
When gradients of the hydrodynamic variables can be neglected, the resulting stress tensor and $U(1)$ current of such a system take the form
\begin{align}
\begin{split}
\label{E:twofluid}
	T^{\nu\sigma} &= \left(\epsilon+P\right)u^{\nu}u^{\sigma} +  P \eta^{\nu\sigma} + \mu \rho_{\rm s} v^{\nu} v^{\sigma}\ ,  \\
	J^{\nu} &= \rho_{\rm n} u^{\nu} + \rho_{\rm s} v^{\nu} \ 
\end{split}
\end{align}
where $\epsilon$ is defined as the Legendre transform of the pressure with respect to the temperature and chemical potential, $\epsilon = - P + Ts+\rho_{\rm n}\mu$. With this definition, $\epsilon$ differs from the time-time component of the energy-momentum tensor. In what follows we will use $w = \epsilon + P$ and $\rho = \rho_{\rm s} + \rho_{\rm n}$. In the superfluid literature, there exists an alternate definition of the condensate density via a current-current correlator. That these two definition coincide was shown in ref.\ \cite{Valle:2007xx}. For convenience, we reproduce the result in appendix \ref{S:twopoint}.

The hydrodynamic equations are given by the conservation equations,
\begin{equation}
\label{E:dmTmn}
	\partial_{\nu} T^{\nu \sigma} = 0 \ , \quad 
	\partial_{\nu} J^{\nu} = 0 \ .
\end{equation}
These conservation conditions are supplemented by a Josephson equation:
\begin{equation}
\label{E:Josephson}
	u^{\nu} v_{\nu} =-1\ .
\end{equation}
One way of deriving the Josephson condition along with the conservation laws
(\ref{E:dmTmn}) and eqs. \eqref{E:twofluid} is the Poisson bracket technique used in refs.\ \cite{Son:2000ht,Valle:2007xx}.
Note that with the absence of dissipative terms in eqs.\ \eqref{E:twofluid}, 
eqs.\ \eqref{E:dmTmn} and eq.\ \eqref{E:Josephson} imply a conserved entropy current:
\begin{equation}
	\partial_{\nu} \left( u^{\nu} s \right) = 0 \ .
\end{equation}

\subsection{Second sound}
\label{S:Secondsound}
To look for sound modes which propagate in the system described by eqs.\ \eqref{E:dmTmn} and \eqref{E:Josephson} one should study linear perturbations of the hydrodynamic variables around a static configuration. 
The system \eqref{E:dmTmn}, \eqref{E:Josephson} supports two sound modes which are given, in the ideal limit where damping effects are ignored, by the two positive roots of
\begin{equation}
\label{E:qubic}
	\alpha c^4 - \beta c^2 + \gamma = 0
\end{equation}
with
\begin{align}
\begin{split}
\label{E:abc}
	\alpha &= T w \left[ \left(\frac{\partial s}{\partial T}\right)_{\mu} \left(\frac{\partial \rho}{\partial \mu}\right)_T - \left(\frac{\partial s}{\partial \mu}\right)_T \left(\frac{\partial \rho}{\partial T}\right)_\mu\right]
	\ ,  
	\qquad
	\gamma  = \frac{s^2 T \rho_{\rm s}}{\mu} \ ,	
	\\
	\beta &= \left(\frac{\partial s}{\partial T}\right)_{\mu} T \left(\rho_{\rm n}^2 + w \frac{\rho_{\rm s}}{\mu}\right) + \left(\frac{\partial \rho}{\partial \mu} \right)_T s^2 T - \left[ \left(\frac{\partial s}{\partial \mu}\right)_T + \left(\frac{\partial \rho}{\partial T}\right)_\mu\right]  s T \rho_{\rm n} \ . 
\end{split}
\end{align}
(We follow ref.\ \cite{Herzog:2008he} in our treatment of the sound modes.)
If the system is conformal so that the only dimensionless quantity is $\mu/T$, then 
one of the two positive roots of eq.\ \eqref{E:qubic} is given by
\begin{equation}
\label{E:sound1}
	c_1^2 = \frac{1}{d}.
\end{equation}
The mode associated with this phase velocity is called normal sound.
The 
other of the two roots of eq.\ \eqref{E:qubic} is then given by
\begin{equation}
\label{E:sound2V1}
	c_2^2 = \frac{s T \rho_s}{w \mu \rho} 
\left( \frac{1}{\rho} \left(\frac{\partial \rho}{\partial \mu}\right)_T  -  \frac{1}{s} \left(\frac{\partial s}{\partial \mu } \right)_T \right)^{-1} \ 
\end{equation}
and is called second sound. 
If instead of the entropy per unit volume $s$ we use the entropy per particle $\sigma = s/\rho$, 
then eq.\ \eqref{E:sound2V1} reduces to
\begin{equation}
\label{E:sound2V2}
	c_2^2 = \frac{\sigma^2 \rho_s}{w} \frac{1}{(\partial \sigma/ \partial T)_\mu} \ ,
\end{equation}
where we have used $(\partial \sigma / \partial \mu)_T = -(T/\mu) (\partial \sigma/ \partial T)_\mu$ which follows from scale invariance. 

The relativistic formulation of hydrodynamics developed in ref.\ \cite{Son:2000ht} and described above coincides with the relativistic version of the Landau-Tisza two fluid model for superfluids \cite{KhalatnikovRelativistic,Carter:1993aq}. In what follows we would like to compare eq.\ \eqref{E:sound2V2} with the non-relativistic expression for second sound obtained directly from the Landau-Tisza two fluid model \cite{Tisza,Landau} (see for example ref.\ \cite{Khalatnikov} for a review). In the non relativistic setup first and second sound are given by the solutions to eq.\ \eqref{E:qubic} with
 \begin{align}
\begin{split}
 \label{Khal}
 	\alpha&=1 \ , \qquad
	\beta= \frac{1}{m} \left[
\left( \frac{\partial P}{\partial \rhom} \right)_\sigma + \frac{\rhoms}{\rhomn} \sigma^2 \left(
\frac{\partial T}{\partial \sigma} \right)_{\rhom} \right] \ ,  \\
	\gamma&= \frac{1}{m^2} \frac{\rhoms}{\rhomn} \sigma^2 \left( \frac{\partial T}{\partial \sigma} \right)_{\rhom}
\left( \frac{\partial P}{\partial \rhom} \right)_T \ ,
\end{split}
\end{align}
where $m$ is the mass of the condensing particles and $\rhom$, $\rhoms$ and $\rhomn$ are the total  number density, superfluid number density, and normal component number density respectively.
The quadratic equation is typically solved by making an additional assumption about the compressibility of the fluid, $\kappa_T = \kappa_\sigma$ where  $\kappa_\sigma \equiv (\partial P / \partial \tilde \rho)_\sigma$ and 
 $\kappa_T \equiv (\partial P /  \partial \tilde \rho)_T$
  The difference $\kappa_T - \kappa_\sigma$
   is proportional to the square of the thermal compressibility of the material.  
   For superfluid ${}^4$He, the thermal compressibility is tiny 
   ($1 - \kappa_\sigma / \kappa_T \sim 10^{-3}$),
 justifying this assumption and the use of the term ``incompressible''.
With this assumption, the non-relativistic result for second sound is
\be
\label{E:nrsound2}
\tilde c_2^2 = \frac{\sigma^2 \rhoms}{m \rhomn} 
\frac{1}{(\partial \sigma / \partial T)_{\rhom}} \ .
\ee

We wish to compare the incompressible result, eq.\ (\ref{E:nrsound2}), with the scale invariant result, eq.\ (\ref{E:sound2V2}).
To that end, 
we define the non-relativistic limit by taking the chemical potential to be very large and approximately equal to the mass of the particles, and the number densities approximately equal to the corresponding charge densities:
\begin{equation}
\label{E:nridentities}
\mu \approx m \ , \qquad 
	w \approx \mu \rho_{\rm n} \ ,
	\qquad
	\rhomn \approx \rho_{\rm n} \ , \qquad \rhoms \approx \rho_{\rm s} \ .
\end{equation}
In this limit, indeed the scale  invariant result (\ref{E:sound2V2}) is close to the incompressible, non-relativistic result. The analysis of ref.\ \cite{Fouxon:2008tb} anticipated such a connection. 

By directly comparing the coefficients  \eqref{Khal} and \eqref{E:abc}, we can recover the non-relativistic sound speeds from the relativistic result in a more general context that does not assume scale invariance or incompressibility.
By replacing the dependent variables $\mu$ and $T$ in the coefficients \eqref{E:abc} with
$\sigma$ and $\rho$, we find, not assuming scale invariance, that
\begin{align}
\begin{split}
\label{E:abcp}
\alpha &= 1 \ , \qquad
\gamma = \frac{1}{\mu}\frac{\rho_{\rm s}}{w} \sigma^2 
\left( \frac{dT}{d \sigma} \right)_\rho \left( \frac{\partial P}{\partial \rho} \right)_T
 \ , \\
\beta &= \frac{\rho_{\rm n}}{w} \left[
\left( \frac{\partial P}{\partial \rho} \right)_\sigma
\left( 1+ \frac{\rho_{\rm s}}{\rho_{\rm n}} \frac{\sigma T}{\mu} \right) - 2 \left( \frac{\partial P}{\partial \sigma} \right)_\rho
\frac{\rho_{\rm s} \sigma^2 T}{ \rho_{\rm n} \rho \mu} 
+ \left( \frac{\partial T}{\partial \sigma} \right)_\rho \sigma^2 \frac{\rho_{\rm s}}{\rho_{\rm n}} \left( 1 - \frac{\sigma T}{\mu} \right) 
\right] 
\ , 
\end{split}
\end{align}
Comparing the coefficients  (\ref{Khal}) with the coefficients \eqref{E:abcp}, 
the corresponding quadratic equations \eqref{E:qubic} agree 
in the norelativistic limit \eqref{E:nridentities}.

For the numerical work in section \ref{S:Numerics}, instead of working with $\sigma$ and $\rho$ as hydrodynamic variables, we find it convenient to use the chemical potential and entropy. To convert from the $T$ and $\mu$ variables of \eqref{E:abc} to the $s$ and $\mu$ variables, we note that in the scale invariant case
\begin{align}
\begin{split}
\label{E:thermoderivatives}
	\left( \frac{\partial \rho}{\partial T} \right)_\mu &= \frac{\rho d - \mu \rho'}{T - T' \mu} \ , \quad 
	\left( \frac{\partial \rho}{\partial \mu} \right)_T = \frac{T \rho' - T' \rho d}{T - T' \mu} \ , \\
	\left( \frac{\partial s}{\partial T} \right)_\mu &= \frac{s d }{T - T' \mu} \ , \quad
	\left( \frac{\partial s}{\partial \mu} \right)_T = -\frac{sT' d}{T - T' \mu} \ , 
\end{split}
\end{align}
where $T' \equiv (\partial T / \partial \mu)_s$ and $\rho' \equiv (\partial \rho / \partial \mu)_s$.
Equality of mixed partial derivatives implies that $(\partial s / \partial \mu)_T = (\partial \rho / \partial T)_\mu$ and hence that
\be
\label{E:Maxwell}
- s T' d + \mu \rho' = \rho d \ .
\ee
Given these expressions, it is easy to see that
\be
\label{E:sound2V3}
c_2^2 = \frac{s \rho_{\rm s}}{w \mu \rho'} \left( T - T' \mu \right) \ .
\ee
Near the critical temperature, second sound vanishes due to the vanishing of the superfluid density $\rho_{\rm s}$.

In non-relativistic superfluid ${}^4$He, it is expected that at low temperatures second sound will approach the speed of sound in a phonon gas.\footnote{%
 If the sample is not pure, i.e.\ $\rho_{\rm s}$ does not approach $\rho$ as $T \to 0$, 
 the second sound speed should vanish at low temperature \cite{Pomeranchuk, Kummer}.
}
This expectation is based on the assumption that at low temperatures the dominant excitations are phonons whose dispersion relation is given by $\omega = k c_q$. 
In ref.\ \cite{Carter:1995if} it was argued that in  a relativistic setting such a dispersion relation will lead to 
\be
\label{E:vpdef}
c_q^2 = \frac{sT}{s T+\mu \rho_{\rm n}}
\ee
and to $T \left(\partial s/ \partial T\right)_{\mu} = s d$. 
Implicit in the identification (\ref{E:vpdef}) is the assumption that only the phonons contribute to $s$ and $\rho_{\rm n}$ at low temperature.  Using eqs.\ \eqref{E:thermoderivatives} and \eqref{E:sound2V3}, and assuming that $\rho \sim \mu^d$ at low temperature, we find
\begin{equation}
	\lim_{T \to 0} c_2^2 = \frac{c_q^2}{d} \ .
\end{equation}
This expression remains unchanged in a non-relativistic setting. See for example  ref.\ \cite{LandLV9}.

\subsection{Fourth sound}
\label{S:Fourthsound}

If the normal component of the fluid is prevented from moving, it is possible to excite a different kind of sound mode, fourth sound \cite{PhysRev.73.608,Atkins}. To observe fourth sound experimentally,
superfluid is channeled through a tube packed with a powder that immobilizes the normal component \cite{Shapiro}.
In such a setup momentum is not conserved; when computing the sound velocity we need, once again, to find the phase velocity for linearized fluctuations around a static background, but here omit the momentum conservation equation \eqref{E:dmTmn}. In the formulation we are using, ref.\ \cite{Yarom:2009uq} carried out such an analysis and found that the phase velocity associated with fourth sound is 
\begin{equation}
\label{E:sound4}
	c_4^2 = \frac{\rho_{\rm s}}{\mu \left(\frac{\partial \rho}{\partial \mu}\right)_s}.
\end{equation}
Ref.\ \cite{Herzog:2008he} observed that fourth sound asymptotes to second sound 
in a large temperature limit.  
To make this relation more transparent, 
we use eq.\ \eqref{E:sound4} to rewrite eq.\ \eqref{E:sound2V2} in the form
\begin{equation}
\label{E:sound4V2}
	c_4^2 = c_2^2 \frac{w}{s^2 d} \left( \frac{ \partial s}{\partial T} \right)_\mu \ .
\end{equation}
At high temperatures, we expect $s \sim T^d$ and $w \approx sT$.

The non-relativistic expression for fourth sound, computed in ref.\ \cite{Atkins}, is given by 
\begin{equation}
\label{E:nrsound4}
	\tilde{c}_4^2 = \frac{\rhoms}{\rhom}\tilde{c}_1^2 + \frac{\rhomn}{\rhom} \tilde{c}_2^2.
\end{equation}
To compare eq.\ \eqref{E:nrsound4} with eq.\ \eqref{E:sound4}, we use eq.\ \eqref{E:Maxwell} and eq.\ \eqref{E:sound2V3} to obtain $\rho^{\prime}$ in terms of $c_2^2$. We can then 
rewrite eq.\ \eqref{E:sound4} as
\begin{equation}
\label{E:sound4V3}
	c_4^2 = \frac{\mu \rho_s}{s T + \mu \rho} c_1^2 + \frac{w}{s T + \mu \rho}c_2^2 \,.
\end{equation}
In the non relativistic limit \eqref{E:nridentities}, the expression \eqref{E:sound4V3} reduces to
 eq.\ \eqref{E:nrsound4}.

Ref.\ \cite{Shapiro} observed experimentally that in ${}^4$He, fourth sound closely follows eq.\ \eqref{E:nrsound4}: 
$c_4^2$
vanishes at the critical temperature, and approaches first sound at low temperatures. A similar behavior is expected from the relativistic formula \eqref{E:sound4V3}. 
At low temperature
and for a pure sample for which $\rho_{\rm s} \sim \rho$ we obtain $c_4^2 \sim c_1^2$.

\section{The Gravity Dual of a superfluid}
\label{S:Bulk}

In this section, with an aim of calculating sound speeds in mind,
we consider 
 a holographic model of a strongly interacting, relativistic, scale invariant superfluid
\cite{Gubser:2008px, Hartnoll:2008vx,Herzog:2008he}.  
The action for a Maxwell field and a charged complex scalar field coupled to gravity is
\be
S = S_{\rm bulk} + S_{\rm boundary}
\label{S}
\ee
where
\be
S_{\rm bulk} =  \int d^5x \, \sqrt{-g} \left[
\frac{1}{2 \kappa_5^2} \left( R + \frac{12}{L^2} \right) - \frac{1}{4e^2} F^{ab}F_{ab} - V(|\psi|) - | \partial \psi - i q A \psi |^2 \right]\ ,
\label{Sbulk}
\ee
with 
\be
\label{E:potential}
V(|\psi|) = m^2 |\psi|^2 + \frac{u^2}{2} |\psi|^4 \ 
\ee
and $S_{\rm boundary}$ is a boundary action which ensures a well posed variational problem and also renders the on-shell action finite.  
We reveal the various terms in $S_{\rm boundary}$ below as we need them. 
Unlike a traditional four dimensional Landau-Ginzburg model, a $u\neq 0$ is not necessary to see a superfluid phase transition in this gravitational system. However, since the zero temperature description of the model with $u \neq 0$ seems to be under good control \cite{Gubser:2008wz}, we will be interested in seeing the effect of a higher order term on the low temperature behavior of the system.\footnote{%
In \cite{Franco:2009yz} a different bulk action was considered where a global $U(1)$ symmetry is broken via the Stuckelberg mechanism. It would be interesting to study the behavior of second sound in such a setup.
}
Roman indices $a, b, \ldots$ run from $0$ to $4$ and are raised and lowered with the five dimensional metric $g_{ab}$. The Greek indices $\mu, \nu, \ldots$ used in the discussion of sound speeds, and which run from $0$ to $3$, are raised and lowered with the Minkowski tensor $\eta^{\mu\nu}$ with signature $(-+++)$.

We look for solutions to the equations of motion that follow from this action that are asymptotically anti-de Sitter with a flat slicing.  More specifically, we require the metric to approach the form
\be
ds^2 = \frac{r^2}{L^2} (-dt^2 + d x_i^2) + L^2 \frac{dr^2}{r^2}
\ee
at large $r$. The conformal boundary of the space is the constant $r$ slice at $r \to \infty$.
From the usual rules of the AdS/CFT duality \cite{Gubser:1998bc,Witten:1998qj} the boundary value of the metric $g_{\mu\nu}$ acts as a source for the boundary theory stress tensor $T^{\mu\nu}$; the bulk gauge field $A_\mu$ acts as a source for a conserved  current $J^\mu$ corresponding to a global U(1) symmetry; and the near boundary data of the scalar $\psi$ sources a scalar operator $O_\Delta$ with conformal scaling dimension $\Delta \geq 1$. The conformal dimension of $O_\Delta$ is related to the five dimensional mass of the scalar field $m^2$ through $m^2 L^2 = \Delta (\Delta - 4)$.  In what follows, we will be considering the particular case of $m^2 = -15 / 4 L^2$.  This mass yields operators of dimension of either 3/2 or 5/2 depending on which term in the near boundary series expansion of $\psi$ we choose to be our source \cite{Klebanov:1999tb}.
By boundary values, we mean more specifically 
\be
\label{E:boundarysources}
g_{\mu\nu}^{(b)} = \lim_{r \to \infty} (L/r)^{2} g_{\mu\nu} \ , \; \; \;
A_\mu^{(b)} = \lim_{r \to \infty} A_\mu \ , \; \; \;
\psi^{(b)} = \lim_{r \to \infty} (r/L^2)^{4-\Delta} (\psi L^{3/2} )\ .
\ee
Given these definitions, the one-point functions of the corresponding field theory operators are
\be
\langle T^{\mu\nu} \rangle = 
\lim_{r \to \infty} \frac{2}{\sqrt{-g^{(b)}}} \frac{\delta S}{\delta g_{\mu\nu}^{(b)}} \ , \; \; \;
\langle J^\mu \rangle = \lim_{r \to \infty} \frac{1}{\sqrt{-g^{(b)}}} \frac{\delta S}{\delta A^{(b)}_\mu} \ , \; \; \;
\langle O_\Delta \rangle = \lim_{r \to \infty} \frac{1}{\sqrt{-g^{(b)}}} \frac{\delta S}{\delta \psi^{(b)}} \ .
\label{onepoint}
\ee
We will see shortly that $\langle T^{\mu\nu} \rangle$,  $\langle J^{\mu} \rangle$, and $\langle O_{\Delta} \rangle$ are proportional to the coefficients of the $\mathcal{O}(r^{-2})$,  $\mathcal{O}(r^{-2})$, and $\mathcal{O}(r^{-\Delta})$ terms in a near boundary series expansion of the metric $g_{\mu\nu}$, gauge field $A_{\mu}$, and scalar field respectively.

The utility of this action is that it describes a system that undergoes a superfluid phase transition.
If the bulk scalar field vanishes then the equations of motion following from eq.\ \eqref{Sbulk} admit a Reissner-Nordstrom black hole solution. Such a gravity solution corresponds to a thermal phase of the field theory with a nonzero chemical potential. Following the general arguments in ref.\ \cite{Gubser:2008px} once the temperature of the black hole falls below a certain critical value $T_c$, a new phase exists in which the scalar field condenses. 
A solution with a nontrivial profile for $\psi$ in the bulk 
corresponds to a superfluid phase in the field theory. In what follows we will construct such solutions and map bulk to boundary quantities explicitly.
We first describe the construction of a static and stationary configuration of the superfluid which is similar to the solutions constructed in refs.\ \cite{Hartnoll:2008kx,Gubser:2008pf}. In section \ref{S:linearsolution} we study linear vector perturbations of this configuration which will enable us to obtain the remaining thermodynamic quantities required to compute the various sound modes.

\subsection{Static and isotropic configurations}
\label{S:bgdsolution}

First, we construct static and isotropic configurations.
Such configurations are described by the metric and gauge field
\be
\label{E:ansatz1}
	ds^2 = -f(r) dt^2 +  \frac{dr^2}{g(r)} + \frac{r^2}{L^2} d  x_i^2\ ,
	\qquad
	A = \phi(r) dt \ 
\ee
where $f(r)$ and $g(r)$ have a simple zero at $r=r_0$ which is a black hole horizon.
We make the gauge choice that $\psi(r)$ is a real function.  
The equations of motion for $\psi$, $\phi$, $f$, and $g$ are
\begin{align}
\begin{split}
\label{E:EOMbgrd}
0&= \psi'' + \left( \frac{g'}{2g} + \frac{f'}{2f} + \frac{3}{r}  \right) \psi'
+ \frac{q^2 \phi^2}{fg} \psi - \frac{V'}{2g} \ , \\
0 &= \phi'' + \left( \frac{g'}{2g} - \frac{f'}{2f} + \frac{3}{r} \right) \phi' 
- \frac{2 e^2 q^2 \psi^2}{g}\phi \ , \\
0 &= (\psi')^2 + \frac{3 g'}{4 \kappa_5^2 r g } - \frac{3f'}{4 \kappa_5^2 r f} + 
\frac{q^2 \phi^2 \psi^2}{fg} \ , \\
0 &= \frac{(\phi')^2}{2e^2 f} + \frac{3g'}{4 \kappa_5^2 r g} + \frac{3 f'}{4 \kappa_5^2 r f} + \frac{V}{g} - \frac{6}{L^2 \kappa_5^2 g} + \frac{3}{\kappa_5^2 r^2} \ .
\end{split}
\end{align}

Near the boundary we require that $f = r^2/L^2 + \mathcal{O}(r^1)$, $\phi = \mu + \mathcal{O}(r^{-1})$ with $\mu$ a constant and that the coefficient of the $\mathcal{O}(r^{\Delta-4})$ term in a series expansion of $\psi$ vanishes. The latter condition ensures that the boundary theory scalar field is not sourced. These boundary conditions imply the near boundary series expansions
\begin{align}
\begin{split}
\label{E:expansionbgrd}
\psi &= \left(\frac{L}{r} \right)^{3/2} \left( \psi^{(0)} + \frac{ L^2 }{r}   \psi^{(1)}+ \ldots \right) \ , \\
\phi &= \mu - \frac{ e^2 L^3 \rho}{2r^2} + \ldots \ , \\
f(r) &= \frac{r^2}{L^2} - \frac{2  \kappa_5^2 L^3 P}{r^2} + \ldots  \ , \\
g(r) &= \frac{r^2}{L^2} + \frac{\kappa_5^2 L (\psi^{(0)})^2   }{ r} - 
\frac{  2  \kappa_5^2 L^3P - 5\kappa_5^2 L^3 \psi^{(0)}\psi^{(1)}  / 2}{r^2} + \ldots \ , 
\end{split}
\end{align}
with either $\psi^{(0)}$ or $\psi^{(1)}$ set to zero.  
The coefficients $\rho$, $P$, and $\psi^{(0)}$ or $\psi^{(1)}$ are integration constants which are determined by requiring that $f$, $g$ and $\phi$ vanish at the black hole horizon located at $r=r_0$, and that $\psi$ is finite there. Since the boundary value of the gauge field sources the $U(1)$ current, we may interpret the boundary value of $A_t$ as the chemical potential $\mu$. We have used the letters $P$ and $\rho$ in anticipation that these are the pressure and charge density respectively.

The partition function  $\mathcal{Z}$ for this configuration can be computed by evaluating the on-shell Euclidean action $S_E$.\footnote{
Strictly speaking, Euclidean time should be compactified on a circle with a radius equal to the inverse temperature. For brevity, we avoid writing this out explicitly.}  
The Euclidean action is different from the Lorentzian action given in eq.\ \eqref{S} by a minus sign:
\begin{equation}
	\ln \mathcal{Z} = -S_E = S \ .
\end{equation}
The bulk on-shell action \eqref{Sbulk} reduces to a total derivative. Introducing a UV regulator $r_{\infty}$ we find,
\be
\label{E:SbulkOS}
 S_{\rm bulk} =\left.  -  \frac{1}{2 \kappa_5^2 L^3} \int d^4x \, 2 r^2 \sqrt{ f g} \right|_{r=r_\infty} \ .
\ee
The boundary action in \eqref{S} is composed of a Gibbons-Hawking term which ensures a well posed variational problem for the metric
\be
\label{E:GHterm}
	S_{\rm GH} = \left.  \frac{1}{2 \kappa_5^2} 
	\int d^4x \, \sqrt{-\ginf} \, (2K - 6/L)\right|_{r = r_\infty} \ .
\ee
In eq.\ \eqref{E:GHterm}, $\ginf$ is the induced metric on the large radius slice $r=r_\infty$ and $K = \gamma^{ab} \nabla_a n_b$ is the trace of the extrinsic curvature. ($n^a$ is the outward pointing unit normal vector to the boundary.) Additional terms in $S_{\rm boundary}$ are required in order for the boundary terms associated with the variation of the scalar field to vanish on shell. Thus, we have
\begin{align}
\begin{split}
S_{3/2} &= \left. \int d^4 x \, \sqrt{-\ginf} \, \left( 2 \psi n^a \partial_a \psi + \frac{3}{2} \psi^2/L \right)
\right|_{r=r_\infty} \ , \\
S_{5/2} &= \left.  \int d^4 x \, \sqrt{-\ginf} \, \left( -\frac{3}{2} \psi^2/L \right)
\right|_{r=r_\infty} \ .
\end{split}
\end{align}

Using the boundary action
\begin{equation}
\label{E:Sboundary}
	S_{\rm boundary} = S_{GH}+S_{\Delta} \ ,
\end{equation}
we find that the partition function is
\begin{equation}
	\ln \mathcal{Z} = P \frac{\mbox{Volume}}{T}
\end{equation}
where we have assumed that either $\psi^{(0)}$ or $\psi^{(1)}$ have been set to zero. As anticipated, the integration constant $P$ in \eqref{E:expansionbgrd} may be interpreted as the pressure.
With the boundary action \eqref{E:Sboundary} we can use standard techniques \cite{Gubser:1998bc,Witten:1998qj,Balasubramanian:1999re,deHaro:2000xn,Bianchi:2001kw,Bianchi:2001de} to compute the one-point functions \eqref{onepoint}. We find that
\begin{align}
\begin{split}
\label{E:Bulktoboundary1}
	\langle T^{\mu\nu} \rangle &= \hbox{diagonal} \begin{pmatrix}
		3 P & P & P & P
	\end{pmatrix} \ ,  \qquad
	\langle J^{\mu} \rangle =  \begin{pmatrix} \rho & 0 & 0 & 0 \end{pmatrix} \ , \\
	\langle O_{3/2} \rangle &= -2 \psi^{(0)}  \ , \qquad
	\langle O_{5/2} \rangle =  2  \psi^{(1)}  \ .	
\end{split}
\end{align}
The details of this computation are left to the appendix.
Apart from the pressure, chemical potential and charge density, other useful thermodynamic quantities are the temperature and entropy.  
The entropy of the configuration is given by the entropy of the black hole which reads
\begin{equation}
\label{E:entropy}
	s = \frac{2 \pi r_0^3}{\kappa_5^2 L^3}.
\end{equation}
In a similar manner, the temperature is given by the Hawking temperature:
\begin{equation}
\label{E:temperature}
	T = \frac{\sqrt{f' g'}}{4\pi}\Big|_{r=r_0}.
\end{equation}
The absolute value of the expectation value of the scalar field $|\langle O_{\Delta} \rangle|$ can be interpreted as the order parameter for the phase transition \cite{Hartnoll:2008vx}.
Thus, with a static and stationary configuration, we are able to determine all the thermodynamic parameters of the superfluid except for the superfluid density. 
We obtained explicit solutions to eqs.\ \eqref{E:EOMbgrd} numerically
and discuss some details of the results in section \ref{S:Numerics}.

\subsection{Vector perturbations of the static and isotropic system}
\label{S:linearsolution}

In order to determine the superfluid density $\rho_{\rm s}$, we consider a configuration where the superfluid velocity $v_{\nu}$ and normal component velocity $u_{\nu}$ are small, unequal and in the `$x$' direction (we define ${\bf x} = (x,\,y,\,z)$). In such a setup the space-time component of the energy momentum tensor and the space component of the gauge field are related to $u_x$, $v_x$ and $\rho_{\rm s}$ through 
\begin{subequations}
\label{E:TandJsimple}
\begin{align}
\label{Txtsimple}
 T^{tx}  &= (\epsilon + P ) u_x + \mu \rho_{\rm s} 
v_x  
= (sT + \mu \rho_{\rm n}) u_x + \mu \rho_{\rm s} v_x 
\ , \\
J^x  &= \rho_{\rm n} u_x + \rho_{\rm s} v_x \ .
\label{Jxsimple}
\end{align}
\end{subequations}
As described in section \ref{sec:soundmodes}, $v_x$ is proportional to the gradient
of the Goldstone boson $\varphi$:
\begin{equation}
\label{E:defxi}
	\xi_x = \mu v_x = \partial_x \varphi.
\end{equation}
Thus, by computing $\langle T^{xt} \rangle$, $\langle J^{x} \rangle$ and $\xi_x$, we can 
use eqs.\ \eqref{Txtsimple}, \eqref{Jxsimple} and \eqref{E:defxi} to obtain $v_x$, $u_x$ and, in particular, $\rho_{\rm s}$.
To generate a solution where $u_x$ and $v_x$ are small, we consider linear fluctuations of $A_x$ and $G_{tx}$ around the background \eqref{E:ansatz1}.
Instead of the background metric and gauge field in \eqref{E:ansatz1} we use
\be
\label{E:ansatz2}
ds^2 = -f(r) dt^2 +  \frac{dr^2}{g(r)} + \frac{r^2}{L^2} d x_i^2 + \frac{2 r^2}{L^2} G_{tx}  dt dx\ ,\qquad
A = \phi(r) dt + A_x(r) dx 
\ee
and work to linear order in $G_{tx}$ and $A_x$. 

The linearized equations of motion for $G_{tx}$ and $A_x$ around the background \eqref{E:ansatz2} are:
\begin{align}
\begin{split}
\label{E:vectormodesEOM}
0&= A_x'' + \left( \frac{f'}{2f} + \frac{g'}{2g} + \frac{1}{r} \right) A_x' - \frac{2 e^2 q^2 \psi^2}{g} A_x
+ \frac{r^2 \phi'}{f} \frac{G_{tx}'}{L^2} \ , \\
0 &= \frac{G_{tx}''}{L^2}+ \left( \frac{g'}{2g} - \frac{f'}{2f} + \frac{5}{r} \right) \frac{G_{tx}'}{L^2}
+ \frac{2  \kappa_5^2}{e^2} \frac{\phi' }{r^2} A_x' + 
\frac{4 \kappa_5^2 q^2 \phi \psi^2 }{r^2 g} A_x \ . 
\end{split}
\end{align}
Note that $A_x=0$ and $G_{tx}=C$, a constant, automatically solve the equations of motion. At the linearized level, this solution is equivalent to a diffeomorphism $x \to x + C t$. From the 
definition \eqref{E:boundarysources} we see that this diffeomorphism also acts on the boundary metric. In order to retain a Minkowski metric at the boundary, we will set the boundary value of $G_{tx}$ to zero. 
Another solution to eq.\ \eqref{E:vectormodesEOM} is 
\begin{equation}
\label{E:Boostsolution}
A_x = \phi C \ , \quad G_{tx} = \left(- f(r)  L^2/r^2+ 1\right)C \ , 
\end{equation}
with $C$ a constant. This solution corresponds to the coordinate transformation $x \to x - C t$, $t \to t-C x$, an infinitesimal boost along the $x$ direction, under which the boundary theory Minkowski metric is invariant. Note that this solution satisfies $A_x(r_0)=0$.
There are two more linearly independent solutions to eq.\ \eqref{E:vectormodesEOM}. One of them diverges at the horizon, located at  $r=r_0$, and we obtain the remaining one numerically in section \ref{S:Numerics}.

With the boundary conditions we have discussed, the near boundary expansions of $G_{tx}$ and $A_x$ take the form
\begin{equation}
\label{E:expansionlnr}
	G_{tx} = -\frac{t  \kappa_5^2 L^5}{2 r^4}  + {\mathcal O}(r^{-5}) \ , \quad
	A_x = -\xi + \frac{j e^2 L^3}{2 r^2} + {\mathcal O}(r^{-3}) \ .
\end{equation}
Recall that we are working in a gauge in which the scalar is real. If we act on 
the series expansion \eqref{E:expansionlnr} with a gauge transformation of the form
\begin{equation}
	A_x \to A_x + \partial_x \left( \xi x \right) \ , \quad 
	\psi \to \psi e^{i q \xi x} \ ,
\end{equation}
we will be in a frame where the boundary value of the gauge field vanishes.
In this frame we can use \eqref{E:defxi} to identify $\xi$ with the gradient of the phase of the scalar. We will see shortly that $j$ is equal to the spatial component of the $U(1)$ current and $t$ gives us the space-time component of the boundary theory stress-energy tensor.

As before, we can compute the partition function for this configuration. Using the on-shell action \eqref{S} with the bulk action \eqref{Sbulk} and boundary action \eqref{E:Sboundary}, we find
\begin{equation}
		\ln \mathcal{Z} =\left(P - \xi j/2 \right) \frac{\mbox{Volume} }{T}
\end{equation}
where we have kept only terms which are quadratic in $G_{tx}$ and $A_x$.\footnote{%
 An intermediate result in this computation is the 
 second order contribution to $S_{\rm bulk}$:
\begin{eqnarray*}
 S_{\rm bulk}^{(2)} &=&  \left. -  \frac{1}{2L} \int d^4x \, \sqrt{fg} \left[
\frac{r}{e^2}  A_x A_x' - \frac{r^4}{2 \kappa_5^2} \left( - \frac{4}{f} + r \frac{f'}{f^2} \right) \frac{G_{tx}^2 }{L^4}
+ \frac{3}{2 \kappa_5^2 } \frac{r^5}{f} \frac{G_{tx} G_{tx}' }{L^4}+ \frac{1}{e^2} \frac{ r^3 \phi'}{ f} A_x \frac{G_{tx}}{L^2}  \right]  \right|_{r=r_\infty}\ . 
\end{eqnarray*}
}
Computing the one-point functions from eq.\ \eqref{onepoint}, we find that eq.\ \eqref{E:Bulktoboundary1} receives corrections
\be
\langle T^{tx} \rangle = t \ , \qquad
\langle J^x \rangle = j \ .
\label{onepointresultstwo}
\ee
See appendix \ref{A:counterterms} for details. Following the strategy presented at the beginning of this section, once we have a solution to eqs.\ \eqref{E:vectormodesEOM}, we can can extract $t$, $j$ and $\xi$ from them and then use eqs.\ \eqref{Txtsimple} and \eqref{Jxsimple} together with $\xi = v_x \mu$ to obtain $\rho_{\rm s}$.

\section{Numerical Results}
\label{S:Numerics}

In order to compute first, second and fourth sound explicitly, we solved eqs.\ \eqref{E:EOMbgrd} and \eqref{E:vectormodesEOM} numerically. To obtain the isotropic solution \eqref{E:EOMbgrd}, we found it convenient to use a shooting algorithm. We specified $g'(r_0)$, $\psi(r_0)$ and $\phi'(r_0)$ and looked for a solution where either the $\mathcal{O}(r^{-3/2})$ term or the $\mathcal{O}(r^{-5/2})$ term in a near boundary series expansion of $\psi$ vanished. Generically, for such a solution one will find $f(r) = f_0 r^2/L^2 + \mathcal{O}(r)$ near the boundary. To obtain the near boundary behavior for $f$ as specified by \eqref{E:expansionbgrd} we rescaled the time coordinate such that $t \to t/\sqrt{f_0}$. This rescaling amounts 
to the shifts $f(r) \to f(r)/f_0$ and $\phi(r) \to \phi(r)/\sqrt{f_0}$.
To solve for  the linear perturbations of $A_x$ and $G_{tx}$ a shooting algorithm is not needed; changing $A(r_0)$ and $G_{tx}(r_0)$ amounts to varying  $v_x$ and $u_x$ but not $\rho_{\rm s}$. 

To read off the various coefficients in \eqref{E:expansionbgrd} and \eqref{E:expansionlnr} we compared our numerics with a near boundary series expansion of the bulk fields expanded to $\mathcal{O}(r^{-16})$. The comparison was carried out on a surface of constant $r= 10^n$ (near the boundary) with $n$ an integer. For each solution we made sure that our fit changed by less than $0.01\%$ under the rescaling $n \to n+1$. To reduce numerical error we used at least thirty digits of working precision in our computations and $n \geq 4$. As the temperature was reduced both $n$ and the working precision were increased in order to retain an accurate result. The main restriction on our numerical computations was CPU time.

\begin{figure}
\begin{center}
\includegraphics[width=6 in]{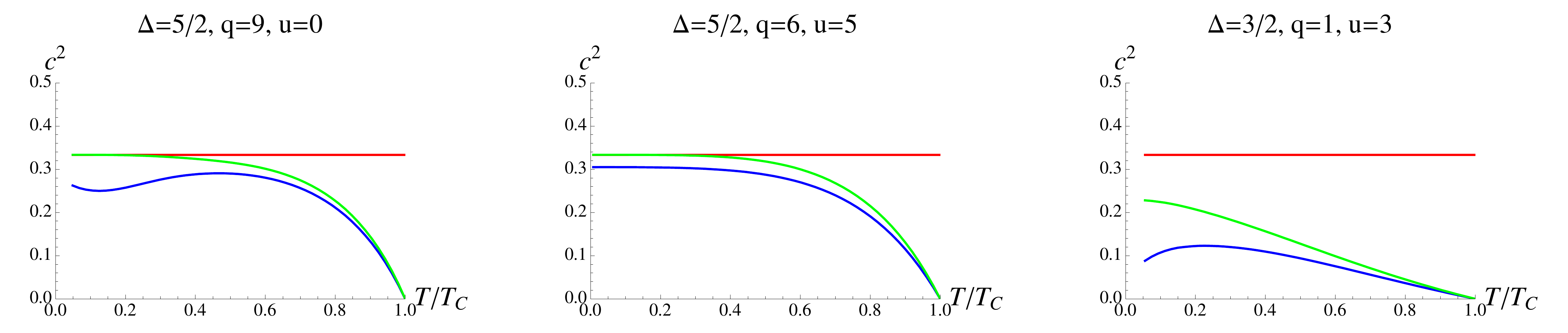}
\caption{\label{F:allthree}
(Color online) First, second and fourth sound for a holographic superfluid given by the bulk action \eqref{Sbulk} and a scalar potential \eqref{E:potential}. From bottom to top, the blue line corresponds to second sound, the green line to fourth sound, and the red line corresponds to first sound computed directly from \eqref{E:abc}.  Note that conformal invariance implies that first sound squared must be $1/3$ for our 3+1 dimensional system.}
\end{center}
\end{figure}

With the near boundary values of $f$, $g$, $\phi$, $\psi$, $G_{tx}$ and $A_x$ in hand, we used eqs.  \eqref{E:entropy}, \eqref{E:temperature} and \eqref{E:expansionlnr} to compute $P$, $\mu$, $\rho$, $\rho_{\rm s}$, $s$ and $T$. We could then compute second and fourth 
sound from eqs.\ \eqref{E:sound2V3} and \eqref{E:sound4}. Some typical results are shown in figure \ref{F:allthree}. 

\begin{figure}
\begin{center}
\includegraphics[width=6 in]{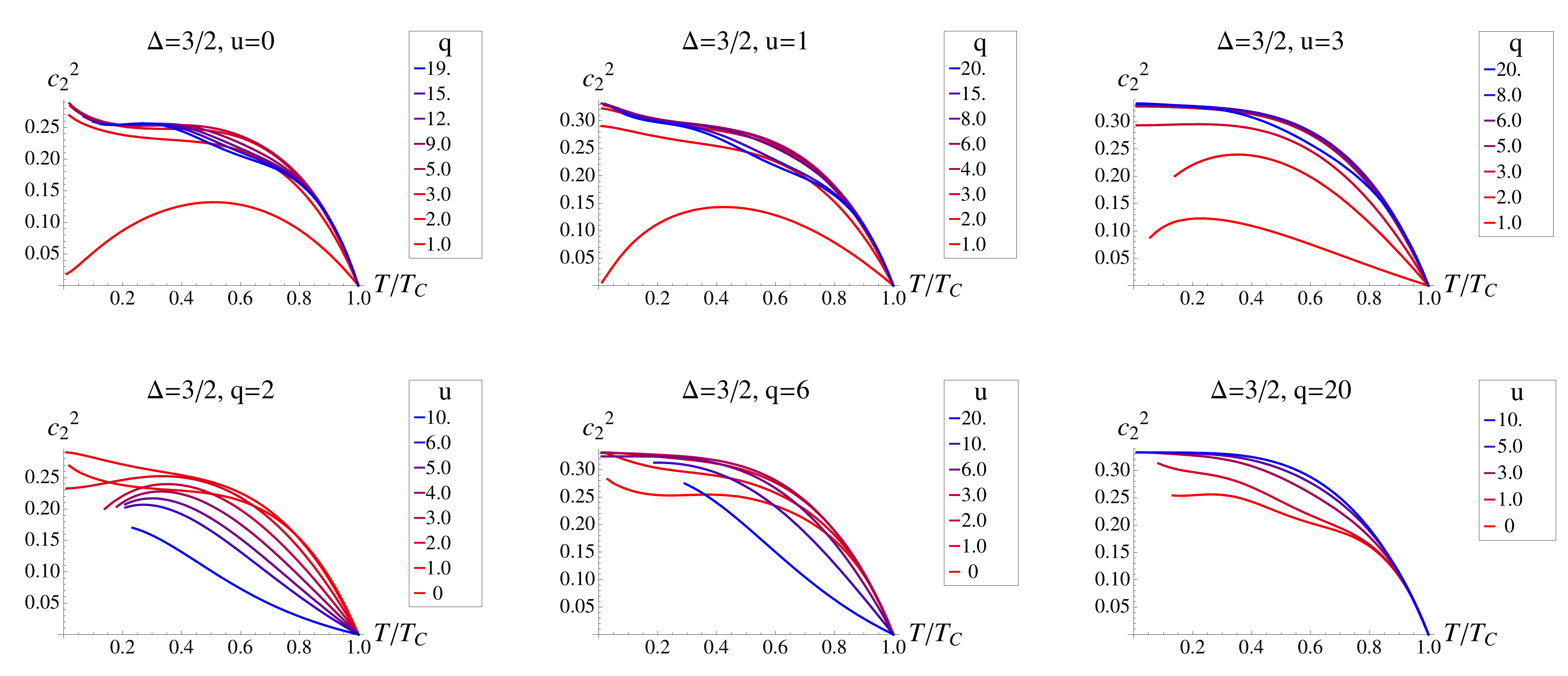}
\caption{\label{F:scalar1}
(Color online) Second sound for a condensate of conformal dimension $3/2$, whose dynamics follow from the bulk action \eqref{Sbulk} and a scalar potential \eqref{E:potential}. }
\end{center}
\end{figure}

\begin{figure}
\begin{center}
\includegraphics[width=6 in]{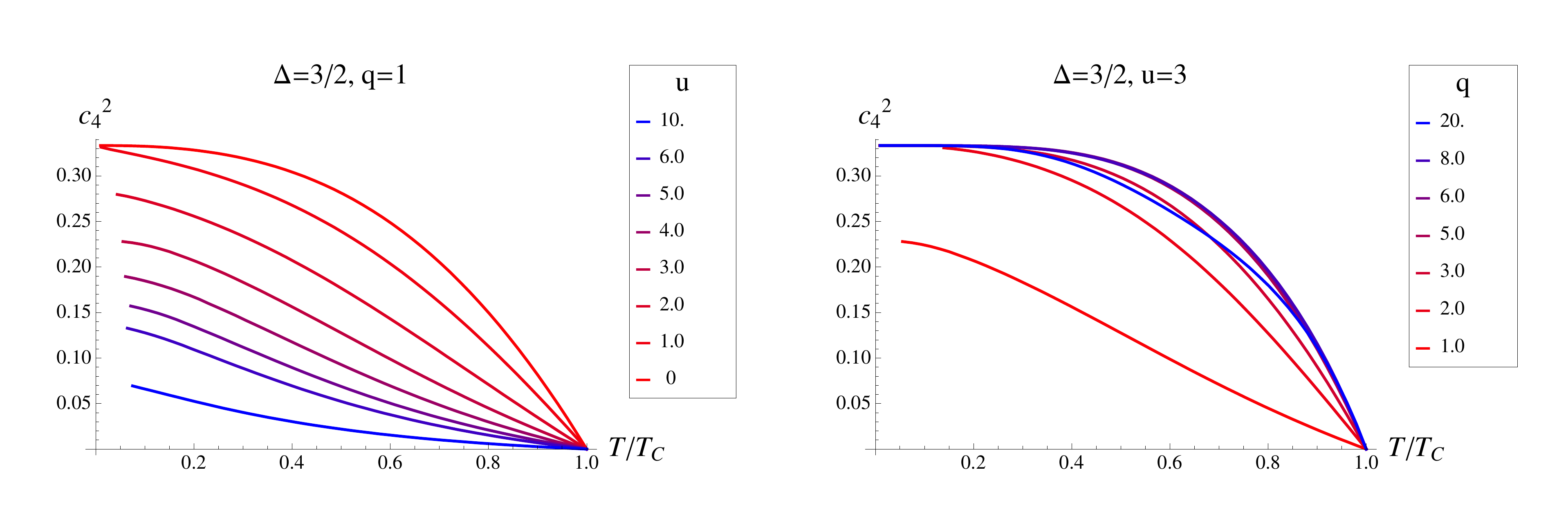}
\caption{\label{F:v4plot}
(Color online) Fourth sound for a condensate of conformal dimension $3/2$ and charge $q=1$, whose dynamics follow from the bulk action \eqref{Sbulk} and a scalar potential \eqref{E:potential}.}
\end{center}
\end{figure}

From figure \ref{F:scalar1} we find that for the $\Delta=3/2$ condensate, there is a critical value of $q$, $q_c(u)$, above which second sound is monotonic. For $q<q_c(u)$, second sound exhibits a maximum. It appears that $q_c(u)$ increases with $u$.
The behavior of fourth sound seems to follow a similar trend: For $q>q_c(u)$ it asymptotes, as expected, to first sound at small temperatures but for $q<q_c(u)$ it falls short of $c_1^2$ at small $T$. A typical example of such behavior can be seen in figure \ref{F:v4plot}.

\begin{figure}
\begin{center}
\includegraphics[width=6 in]{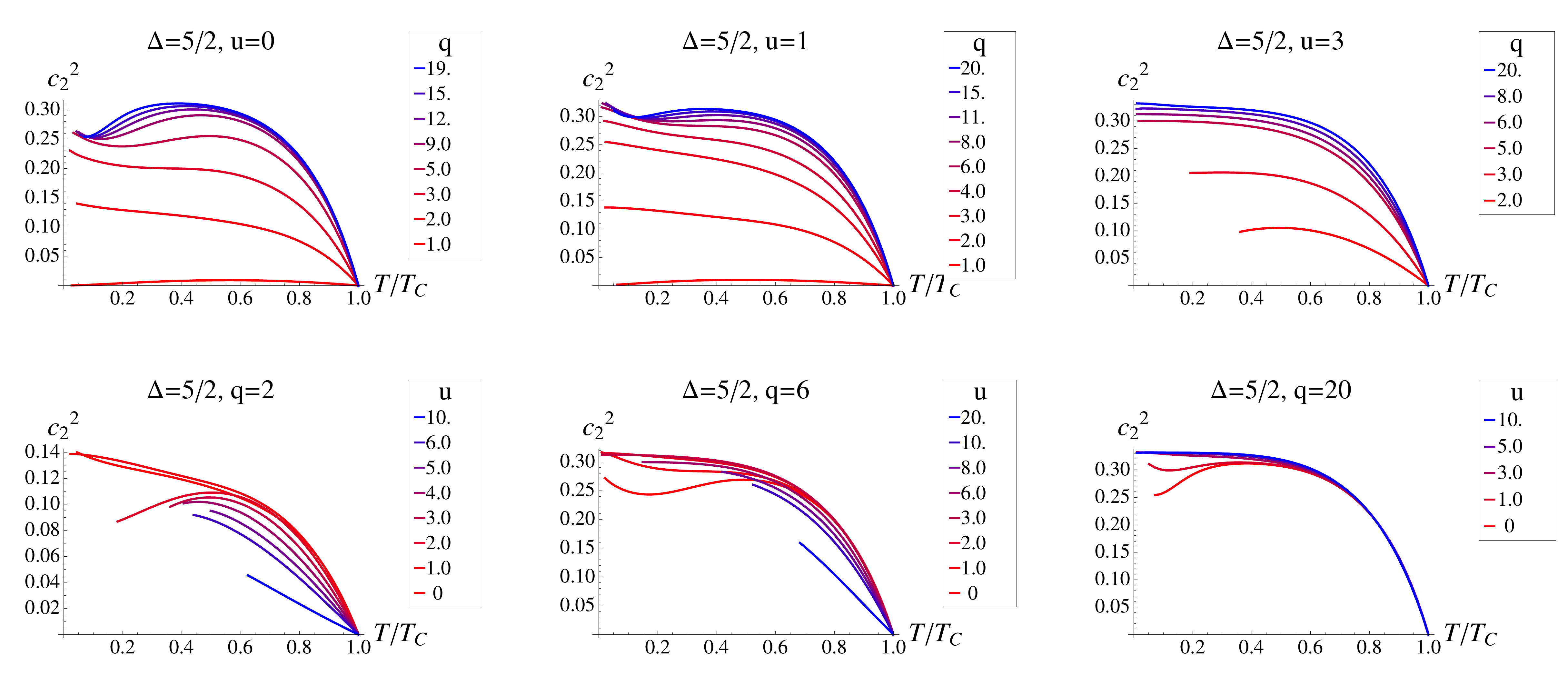}
\caption{\label{F:scalar2}
(Color online) Second sound for a condensate of conformal dimension $5/2$, whose dynamics follow from the bulk action \eqref{Sbulk} and a scalar potential \eqref{E:potential}. }
\end{center}
\end{figure}

The $\Delta = 5/2$ condensate also seems to have a critical charge $q_c(u)$ below which second sound has a distinct maximum. However, for relatively large $q>q_c(u)$ and small $u$, the second sound curves develop a local minimum. This is depicted in figure \ref{F:scalar2}. 
It is tempting to speculate about the possible relevance of roton like excitations in connection with this minimum. The behavior of fourth sound for $\Delta = 5/2$ condensates is similar 
to that of the $\Delta = 3/2$ condensates.

\section{Discussion}
\label{S:Discussion}

In figures \ref{F:scalar1}, \ref{F:v4plot} and \ref{F:scalar2} we presented our main results for second and fourth sound for a holographic superfluid with a bulk action as in \eqref{S}. The prominent features of these results are the appearance of a critical value of the scalar charge $q_c(u)$  and the low temperature limit of second sound which is at odds with Landau's prediction for an incompressible, non relativistic superfluid (see eq. \eqref{E:Landaulimit}).\footnote{%
 The existence of a critical charge was also suggested in ref.\ \cite{Gubser:2008pf}.
}

The appearance of a critical value for the scalar charge may be related to the two mechanisms that can drive the phase transition to the superfluid state \cite{Gubser:2008px,Hartnoll:2008kx,Gubser:2005ih,Denef:2009tp}.
In the first, more obvious mechanism, the coupling of the scalar to the gauge field via the covariant derivative of the scalar generates an effective mass term of the form
\be
m_{\rm eff}^2 = m^2 + q^2 g^{tt} A_t^2 \psi^2 \ .
\ee
Once the charge of the black hole increases beyond a certain critical value, the effective mass of the scalar will become low enough to generate an unstable mode and the scalar will condense.
The second mechanism, noted in the current context in ref.\ \cite{Hartnoll:2008kx}  (but see also \cite{Gubser:2005ih}), is associated with a violation of the Breitenlohner-Freedman bound \cite{Breitenlohner:1982bm,Breitenlohner:1982jf}.
As the temperature is decreased, the black hole approaches extremality.  The near horizon 
metric of an extremal black hole resembles $AdS_2 \times S^2$, and this $AdS_2$ is only stable for scalars above a modified Breitenlohner-Freedman bound. 
Neutral scalars with a mass $m$ satisfying $-d^2/4 \leq m^2 L^2 < -d (d+1) /4$ will condense at low enough temperatures. It is possible that the critical charge we observe is the charge below which the second instability becomes most important in causing 
the phase transition. Another perhaps related possibility, 
investigated in ref.\ \cite{SteveandAbhi}, is that $q_c(u)$ is associated with nonconformal IR behavior of the theory. 

\begin{figure}
\begin{center}
\includegraphics[width=6 in]{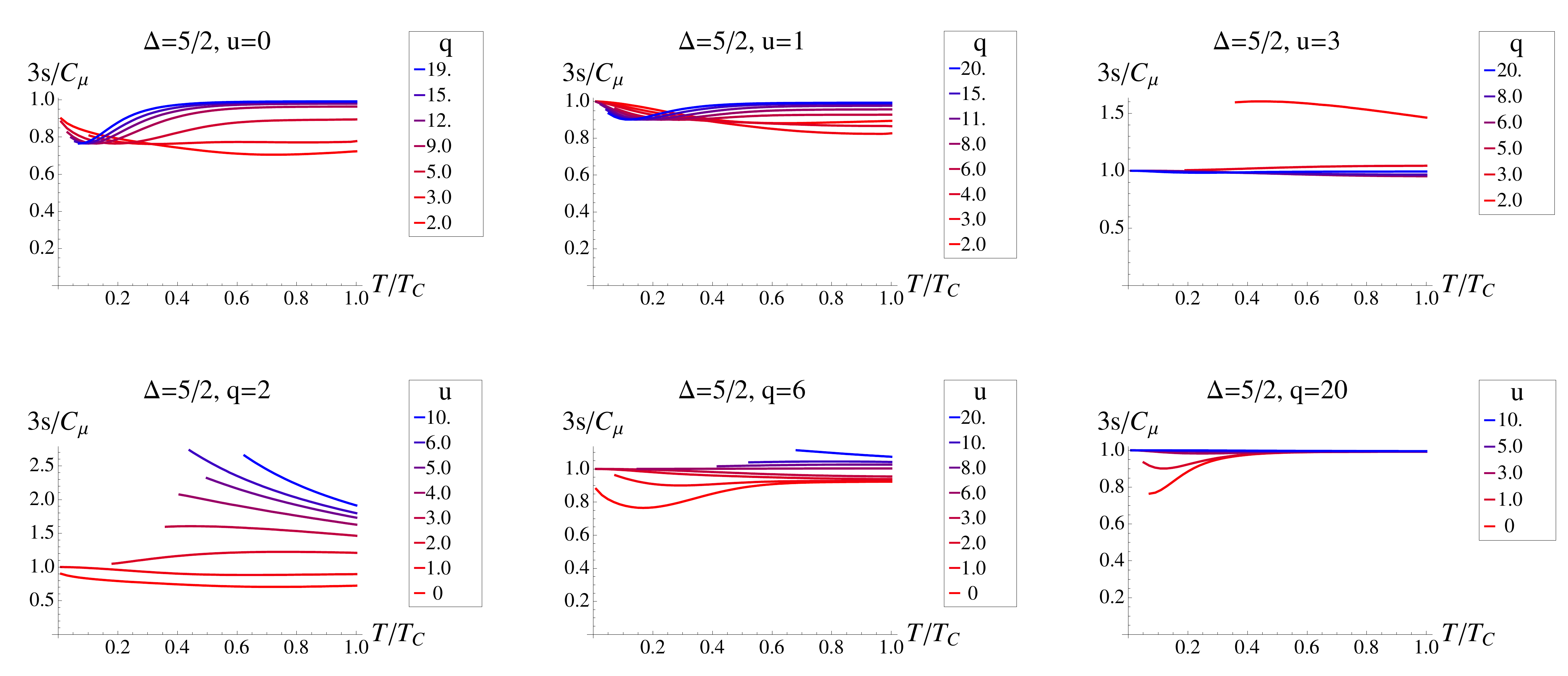}
\caption{\label{F:Cmuplots}
(Color online) Plots of the ratio of the heat capacity at constant chemical potential $C_{\mu}$ to $3 s$ for $\Delta = 5/2$. 
The $q=1$ curves for $u=0$, $u=1$ and $u=3$ have not been displayed --- they are of order 3 with a maximum around $T/T_c \sim 0.1$. The value of $3s/C_{\mu}$ for the $\Delta=3/2$ condensate is very similar to the one in the plot.}
\end{center}
\end{figure}

\begin{figure}
\begin{center}
\includegraphics[width=6 in]{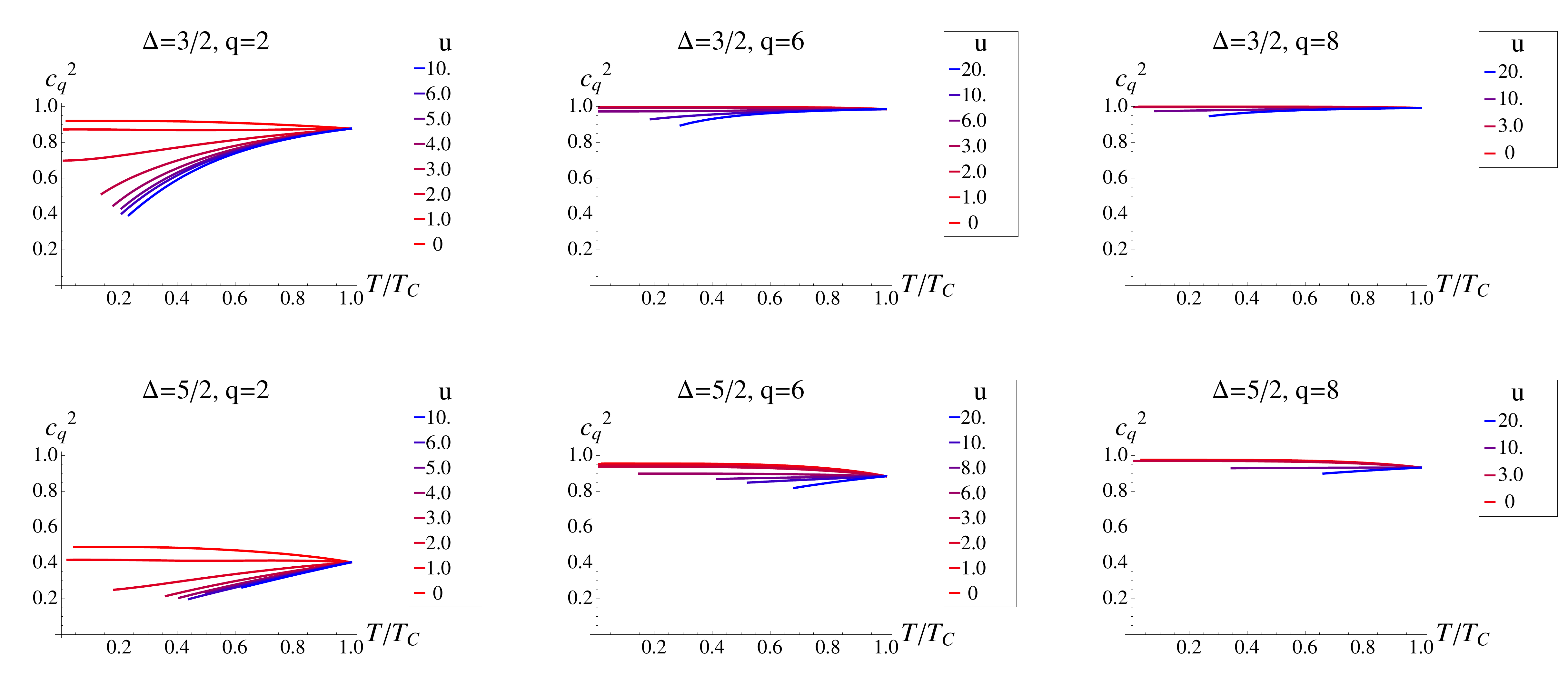}
\caption{\label{F:phonon}
(Color online) Plots of $c_q^2 \equiv s T/(s T+\mu\rho_{\rm n})$. According to \cite{Carter:1995if}, for a relativistic phonon gas, $c_q$ is the phonon velocity. 
}
\end{center}
\end{figure}

In order to understand better the discrepancy between our low temperature results for $c_2^2$ and Landau's prediction, we decompose $c_2^2$ into three pieces:
\begin{equation}
\label{E:sound2V4}
	c_2^2 =  c_4^2 c_q^2  \frac{3s}{C_{\mu}}
\end{equation}
where 
$C_{\mu} \equiv T (\partial s/ \partial T)_{\mu}$ is the heat capacity at constant chemical potential and $c_q^2  \equiv s T/(s T+\mu\rho_{\rm n})$ which, following eq.\ \eqref{E:vpdef}, we call the quasi-particle velocity.
Note that eq.\ (\ref{E:sound2V4}) is a trivial rewriting of eq.\ (\ref{E:sound4V2}).
As discussed in section  \ref{S:Secondsound}, 
if the low lying quasi-particle excitations are phonons, one should find that $C_{\mu}=3s$ and that $c_1=c_q$. 
Moreover, we expect from eq.\ (\ref{E:sound4V3}) that the low temperature limit of fourth sound is first sound.  Instead, as can be seen from figures \ref{F:Cmuplots}, \ref{F:phonon} and \ref{F:v4plot},
in our model
$C_{\mu}$ is not always equal to  $3s$, $c_1$ cannot be identified with $c_q$, and $c_4$ does not always approach $c_1$.
It would be interesting to see if degrees of freedom on the field theory side with a dispersion relation $\omega = c_q k$ can be more directly identified.

We note that when $q$ is very large $C_{\mu}$ approaches $3s$ and  $c_q$ approaches unity for non vanishing temperatures. This behavior is expected; in the probe limit, where the metric does not backreact on the scalar field and gauge field, the spacetime is a neutral black hole in AdS${}_5$ and the chemical potential and density are supressed by a factor of $q^{-1}$ relative to the entropy and temperature.

\begin{figure}
\begin{center}
\includegraphics[width=6 in]{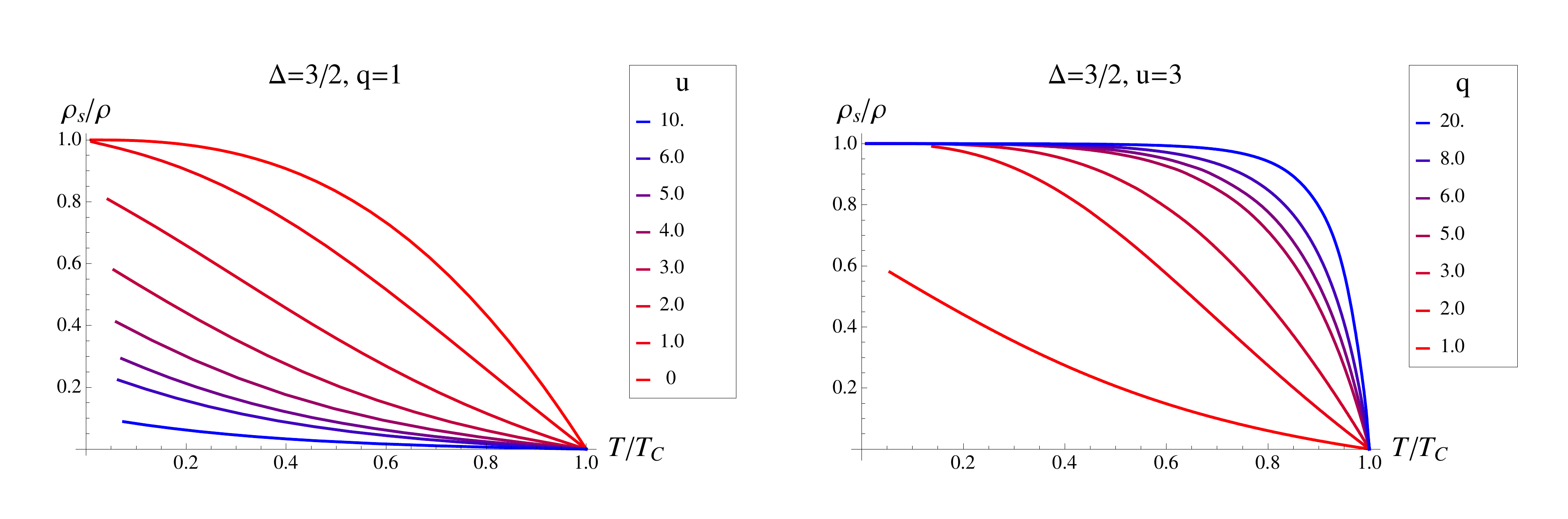}
\caption{\label{F:rhosplot}
(Color online) The relative superfluid density for various condensates. Note the similarity to 
the fourth sound curves depicted in figure \ref{F:v4plot}.}
\end{center}
\end{figure}

The behavior of fourth sound depicted in figure \ref{F:v4plot} can be understood in terms of the behavior of $\rho_{\rm s}$ and $\rho$, c.f. eq. \eqref{E:sound4}. We find that for all values of $q$ and $u$, $\rho \sim \mu^3$ at low temperatures. However, as can be seen from figure \ref{F:rhosplot}, $\rho_{\rm s}$ does not approach $\rho$ for small $q$, at least not for the temperature ranges we've reached. Thus, following the arguments in section \ref{S:Fourthsound} it is not surprising that fourth sound does not asymptote to $1/3$. 
In these cases, second sound seems to vanish at low temperature. 
Similar behavior of second sound has been noted 
in mixtures of ${}^4$He and ${}^3$He where $\rho_{\rm n}$ remains nonzero at low temperatures
\cite{Pomeranchuk, Kummer}.

\section*{Acknowledgments}

We would like to thank P.~Kovtun, S.~Pufu and D.~Son for discussion. CPH would like to thank the Weizmann Institute for hospitality where part of this work was carried out.  The work of CPH was supported in part by the US NSF under Grant No.\ PHY-0756966. The work of AY was supported in part by the DOE under Grant No.\ DE-FG02-91ER40671.

\begin{appendix}

\section{An aside on two-point functions}
\label{S:twopoint}

In section \ref{sec:soundmodes} we defined $\rho_{\rm s}$ as the variable conjugate to $\xi$. This definition is slightly different from the one in the superfluid literature (see for example \cite{NozieresPines}). In what follows we show that these two definitions coincide.

Consider a liquid in which the superfluid component moves with a small velocity $v_x$ while the normal component moves with a small velocity $u_x$.  
In such a configuration, the energy momentum tensor and current are given by \eqref{E:TandJsimple}.
Another way of defining $\rho_{\rm n}$ and $\rho_{\rm s}$ is through a 
zero frequency and long wavelength limit of the Green's functions involving
$T^{xt}$ and $J^x$.
Consider applying a spatially varying external velocity field of the form 
$ U_i e^{i {\bf k x}}$.  
Such a velocity field is canonically conjugate to the momentum density $T^{ti}$, $i=1,\ldots,d$ and we model the effect by adding a term to the Hamiltonian
\be
\label{E:deltaT}
\delta \hat H =  \int d^3{\bf x}  \, T^{ti}(t,{\bf x}) U_i e^{i {\bf k x}} \ .
\ee
Time dependent perturbation theory determines the first order response of the system 
\be
\label{Txtlinear}
\delta \langle T^{tj} \rangle  =  - G_R^{tj,ti}(0, {\bf k}) U_i e^{i {\bf k x}} + \ldots \ ,
\ee
where the Fourier transform of the retarded Green's function is defined as
\be
 G_R^{tj,ti}(k) \equiv i \int d^4x \, e^{-ikx} \theta(t) \langle [ T^{tj}(x), T^{ti}(0) ] \rangle \ .
\ee
By rotational invariance, the Green's function decomposes into two components which give the response of the system to transverse and longitudinal waves
\be
\label{E:Gijdecomposition}
G_R^{ti,tj}(0, {\bf k}) = \frac{k^i k^j}{{\bf k}^2} G^{T\parallel}_{R}({\bf k}) + 
\left( \delta^{ij} - \frac{k^i k^j}{{\bf k}^2} \right) G^{T\perp}_R({\bf k}) \ .
\ee
When ${\bf U}$ is parallel to ${\bf k}$, 
the driving force induces a longitudinal excitation and
we expect both the normal and superfluid components to couple to it.
In the case where ${\bf U}$ is perpendicular to ${\bf k}$ and sufficiently small, 
we expect that no momentum will be transferred to the superfluid component due to its vanishing viscosity. Thus, only
the normal component will be dragged. The transverse wave can be generated by an experiment where an open ended cylinder is filled with superfluid, and is dragged parallel to its own axis.
Using eq.\ (\ref{Txtsimple}) we find that
\be
\lim_{{\bf k} \to 0}  G_R^{T\perp} ({\bf k}) = -(sT + \mu \rho_{\rm n})
\ , \qquad
\lim_{{\bf k} \to 0}  G_R^{T\parallel} ({\bf k}) = -(sT + \mu \rho ) \ .
\ee
These expressions are similar to the corresponding non-relativistic expressions found in the
superfluid literature.  In the non-relativistic limit,  the $sT$ term is dropped because it is negligible compared to $\mu \rho_{\rm n}$ and $\mu \rho$. See eqs.\ \eqref{E:nridentities} in the main text.

A similar analysis can be performed to extract the same limiting behavior of the current-current two-point function.  Instead of a velocity field, we couple the charge current to an external gauge field $A_i = -\mu V_i e^{i {\bf k x}}$. Instead of eq.\ \eqref{E:deltaT} we use
\be
\delta \hat H = \mu \int d^3 {\bf x} \, J^i(t,{\bf x}) V_i e^{i {\bf kx}} \ .
\ee
Performing a decomposition of the retarded current-current two point function into parallel and perpendicular components, similar to eq.\ \eqref{E:Gijdecomposition}, we find
\be
G_R^{i,j}(0, {\bf k}) = \frac{k^i k^j}{{\bf k}^2} G^{J\parallel}_{R}({\bf k}) + 
\left( \delta^{ij} - \frac{k^i k^j}{{\bf k}^2} \right) G^{J\perp}_R({\bf k}) \ .
\ee
If ${\bf V}$ is parallel to ${\bf k}$, then the external field ${\bf A}$ is gauge equivalent to zero, and we
expect no response from the system.  On the other hand, if ${\bf V}$ is perpendicular to ${\bf k}$,
then ${\bf V}$ can be identified with the superfluid velocity from before and we can read off the two-point function from eq.\ (\ref{Jxsimple}),
\be
\lim_{{\bf k} \to 0}  G_R^{J\perp} ({\bf k}) = -\rho_{\rm s} / \mu
\ , \qquad
\lim_{{\bf k} \to 0}  G_R^{J\parallel} ({\bf k}) = 0 \ .
\ee

\section{Counterterms and one-point functions}
\label{A:counterterms}

In this section we compute the one-point functions in \eqref{onepoint} using the general prescription initiated in refs.\ \cite{Gubser:1998bc,Witten:1998qj} 
and elaborated on in refs.\ \cite{Balasubramanian:1999re,deHaro:2000xn,Bianchi:2001kw,Bianchi:2001de}. 

To vary the on-shell action we consider an arbitrary linear perturbation of the solution to the equations of motion.  Under a shift by $\delta A_a$, $\delta \gamma_{ab}$, and $\delta \psi$, the 
action \eqref{S} will shift by an infinitesimal amount.  This amount can be written as a boundary term which we regulate by evaluating it at large $r = r_\infty$.  We find
\begin{subequations}
\label{E:deltaS}
\begin{multline}
	\delta S = -\sqrt{-\gamma} \Bigg( \frac{1}{e^2} n_a F^{ab}  \, \delta A_b
		 +
		 \frac{1}{2 \kappa_5^2} 
		 \left(K^{ab} - K \gamma^{ab} + 3 \gamma^{ab}/L + \frac{3}{4} \psi^2 \ginf^{ab} /L \right) \delta \gamma_{ab} + \\
	+( 2 n^a \partial_a \psi + 3 \psi/L) \delta \psi 
\Bigg) \Bigg|_{r=r_\infty} 
\end{multline}
for $\Delta = 5/2$ and 
\begin{multline}
\delta S = -\sqrt{-\gamma} \Bigg( \frac{1}{e^2} n_a F^{ab}  \, \delta A_b - 2 \psi n^a \partial_a \delta \psi  - 3 \psi \delta \psi/L +
\\
	+ \frac{1}{2 \kappa_5^2} \Bigg(K^{ab} - K \gamma^{ab} + 3 \gamma^{ab}/L 
		+\psi^{} n^{a}\partial^{b}\psi 
		+ \psi^{} n^{b}\partial^{a}\psi 
		\\
		- \left(\psi^{} n^{c}\partial_{c}\psi \right)\ginf^{ab}
		-\frac{3}{4} |\psi|^2 \ginf^{ab}/L \Bigg) \delta \gamma_{ab}
\Bigg) \Bigg|_{r=r_\infty} 
\end{multline}
\end{subequations}
for $\Delta=3/2$. 

Using the background solution \eqref{E:expansionbgrd} and \eqref{E:expansionlnr}, we can express this result as a power series in $1/r$.
For example, examining the coefficient of $\delta A_b$, and working to linear order in $A_x$ and $G_{tx}$, we find that 
\begin{align}
\begin{split}
\sqrt{-\gamma} \, n_a F^{at} &=
 -\rho e^2 
+ {\mathcal O}(r^{-1}) \ , \\
\sqrt{-\gamma} \, n_a F^{ax} &=
-  j e^2 
+ {\mathcal O}(r^{-1}) \ .
\end{split}
\end{align}

Inserting all such expressions into eqs.\ \eqref{E:deltaS} and carrying out the functional differentiation in eqs.\ \eqref{onepoint}, we find, at the end of the day, that 
\begin{align}
\begin{split}
\langle T^{tt} \rangle &= \left( 3P + c_i \psi^{(0)} \psi^{(1)}  \right) 
\ ,
\\
\langle T^{tx} \rangle &= t \ ,
\\
\langle T^{xx} \rangle = 
\langle T^{yy} \rangle  = 
\langle T^{zz} \rangle &= \left(P - c_i  \psi^{(0)} \psi^{(1)} \right)
 \ ,
\\
\langle J^t \rangle = \rho
&\ , \quad
\langle J^x \rangle =  j
\ ,
\\
\langle O_{3/2} \rangle = - 2 \psi^{(0)} 
&\ , \quad
\langle O_{5/2} \rangle = 2 \psi^{(1)} 
 \ ,
\end{split}
\end{align}
with $c_{3/2} = 5/4$ and $c_{5/2} = - 3/4$.
Setting either $\psi^{(0)}$ or $\psi^{(1)}$ to zero,
we recover the results 
 \eqref{E:Bulktoboundary1} and \eqref{onepointresultstwo} in the body of the paper.  The trace of the stress tensor is
 \be
 T^\mu_\mu = -4 c_i \psi^{(0)} \psi^{(1)} =  (4-\Delta) \langle O_\Delta \rangle \psi^{(b)} \ ,
 \ee
in agreement with the conformal Ward identity 
(see for example eq.\ (4.21) of ref.\ \cite{Bianchi:2001kw}).

\end{appendix} 


\bibliographystyle{JHEP}
\bibliography{sound2}

\end{document}